\documentclass[12pt,a4paper]{article}
\usepackage[a4paper, total={6in, 10in}]{geometry}
\usepackage[utf8]{inputenc}
\usepackage{amsmath}
\usepackage{amsfonts}
\usepackage{amssymb}
\usepackage{graphicx}
\usepackage{float}
\usepackage{amsthm}
\usepackage{cancel}
\usepackage{xhfill}
\usepackage{multicol}
\usepackage{xcolor}
\usepackage{soul}
\usepackage{framed}
\usepackage{bm}
\usepackage{subcaption}
\usepackage{CJKutf8}
\usepackage{lineno}
\usepackage[authoryear]{natbib}
\usepackage[
    colorlinks=true,
    linkcolor=blue,
    citecolor=blue,
    urlcolor=blue
]{hyperref}

\title{Exploring a Correlation Between AGN Black Hole Mass and X-ray Luminosity}
\author{Wordsworth K. K. Chu$^{1}$, A. Kathy Romer$^{1}$,  Paul Giles$^{1}$,\\ Jessica E. Pilling$^{1}$, Julian A. Mayers,$^{1}$}
\date{\today}
\begin{document}

\maketitle
\vspace{-1cm}
$^{1}$Astronomy Centre, University of Sussex, Falmer, Brighton, BN1 9QH, UK
\vspace{1cm}
\noindent

\textbf{ABSTRACT}\\

The traditional method to measure Type-1 Active Galactic Nuclei central black hole masses ($M_{BH}$) is the reverberation mapping (RM) technique. Motivated by the desire to find a less observationally expensive method (than RM) to estimate $M_{BH}$, we have explored a purported correlation \citep{Mayers2018} between X-ray luminosity and $M_{BH}$. For this we have compiled a sample of 35 AGN with RM $M_{BH}$ estimates that were detected in the eROSITA All-Sky Survey public data release.
Using a sample that is  larger than that of \citealt{Mayers2018} and has a cleaner selection function, we confirm their XMM-Newton based result of a positive correlation. However, the scatter on the scaling relation remains too high ($\sigma_\mathrm{int}=0.50^{+0.08}_{-0.06}$) to make this a practical method to predict $M_{BH}$ for a given X-ray detected Type-1 AGN. That said, our findings can be used to improve mock X-ray skies \citep{comparat2019active} generated from large volume hydrodynamical cosmological simulations. Such mocks are a necessary ingredient for cluster cosmology \citep{Ghirardini2024} and preparation for future missions such as NewAthena \citep{Covas2025}. 

\section{Introduction and motivation for a new study}

Active Galactic Nuclei (AGNs) are luminous extragalactic sources powered by accretion onto supermassive black holes (SMBHs) at their centres. Their strong X-ray emission provides a valuable probe of the accretion process. Previous studies have demonstrated a strong anti-correlation between X-ray variability and the mass of the central SMBH. This relationship has been explored through analyses of power spectral density (PSD) (e.g. \cite{kording2007}) and normalised excess variance ($\sigma_\mathrm{NXS}^2$) (e.g. \cite{Ponti2011}). However, these methods generally require long duration and high resolution X-ray observations, limiting their applicability to large samples. 

In addition to reporting the correlation between $\sigma^2_\mathrm{NXS}$ and $M_\mathrm{BH}$, \citealt{Ponti2011}, also reported a correlation between $L_\mathrm{X}$ and $\sigma^2_\mathrm{NXS}$. This finding motivated the \citealt{Mayers2018} investigation into a possible direct relationship between $L_\mathrm{X}$ and $M_\mathrm{BH}$. 
 Using XMM-Newton data from the XMM Cluster Survey (XCS) \citep{Romer2001}, they selected 30 Type-1 AGN candidates and measured their X-ray variability, combining these data with $M_\mathrm{BH}$ derived from RM. 

The \citealt{Mayers2018} study was submitted to \textit{Monthly Notices of the Royal Astronomical Society} (MNRAS), but was withdrawn following reviewers' concerns about the methodology. In particular, concerns were raised regarding sample selection, particularly the narrow redshift range ($z=0.005-0.155$), the reliance on the XMM-Newton archive (which has a complex selection function), and the exclusion of sources with negative $\sigma^2_\mathrm{NXS}$. Therefore, a new study, using a different sample is warranted.

\section{A new sample of Type-1 AGN with X-ray luminosities and black hole mass estimates}

To improve on \citealt{Mayers2018}, we have based our study on the eROSITA eRASS:3 catalogue \citep{ramos2026srg}. Unlike the XMM-Newton pointed observations used in \citealt{Mayers2018}, the eROSITA all-sky survey has uniform sensitivity, making it particularly well suited for investigating the statistical properties of X-ray sources. We identified 35 eRASS:3 detected AGNs with RM-based $M_\mathrm{BH}$ estimates in the literature (primarily from \citealt{Bentz2015}, \citealt{Shen2023}). The sample covers a redshift range of $z=0.005-2.660$.
Unlike \citealt{Mayers2018}, we did not exclude sources with negative $\sigma^2_\mathrm{NXS}$ values, as the short duration eRASS:3 observations are not suitable for variability measurements.

X-ray fluxes, and their associated uncertainties, for the 35 AGN were obtained directly from the eROSITA eRASS:3 catalogue \citep{ramos2026srg}. To convert these to luminosities, we adopted a flat $\Lambda$CDM cosmology with $H_0=70\ \mathrm{km\,s^{-1}\,Mpc^{-1}}$, $\Omega_\mathrm{m}=0.27$, and $\Omega_\Lambda=0.73$. 
The luminosity distances ($D_\mathrm{L}$) were calculated from the redshift of the AGNs under this cosmological model. The $L_X$ were then calculated using
\begin{equation}
    L_\mathrm{X}=4\pi D_\mathrm{L}^2 F_\mathrm{X}K(z)
\end{equation}
where $F_\mathrm{X}$ is the X-ray flux  and $K(z)$ is the K-correction.  We used the eRASS3 fluxes in the 0.5–1.0 keV band, which assume a power law spectral model with a photon index of $\Gamma=2$ \citep{ramos2026srg}. Under this assumption, the K-correction factor, $(1+z)^{\Gamma-2}$, is unity, and no additional K-correction was applied. To validate the $L_\mathrm{X}^{0.5-1.0 {\rm keV}}$ values, we compared with those determined by \citealt{Mayers2018}, using more sensitive XMM-Newton data, for the 5 AGN in common between the two samples, and found them to be in good agreement. We note that Section 5.1.2 of \citealt{Mayers2018} includes a thorough examination of the reliability of  AGN $L_\mathrm{X}$ estimates when based on eRASS count rates.

\section{Fitting a scaling relationship between black hole mass and X-ray luminosity}
\label{sec:method}

The relationship between $M_\mathrm{BH}$ and $L_\mathrm{X}$ was fitted using Markov Chain Monte Carlo (MCMC) methods, with the intrinsic scatter and parameter uncertainties estimated from the posterior distributions. Posterior predictive distributions were used to estimate the $1\sigma$ and $2\sigma$ scatter around the best-fitting relation:

\begin{equation}
\log\left(\frac{M_\mathrm{BH}}{10^8 M_\odot}\right)=A+B\log\left(\frac{L_\mathrm{X}}{2\times10^{43}\ \mathrm{erg\,s^{-1}}}\right).
\end{equation}

\begin{figure}
\centering
\includegraphics[width=1\linewidth]{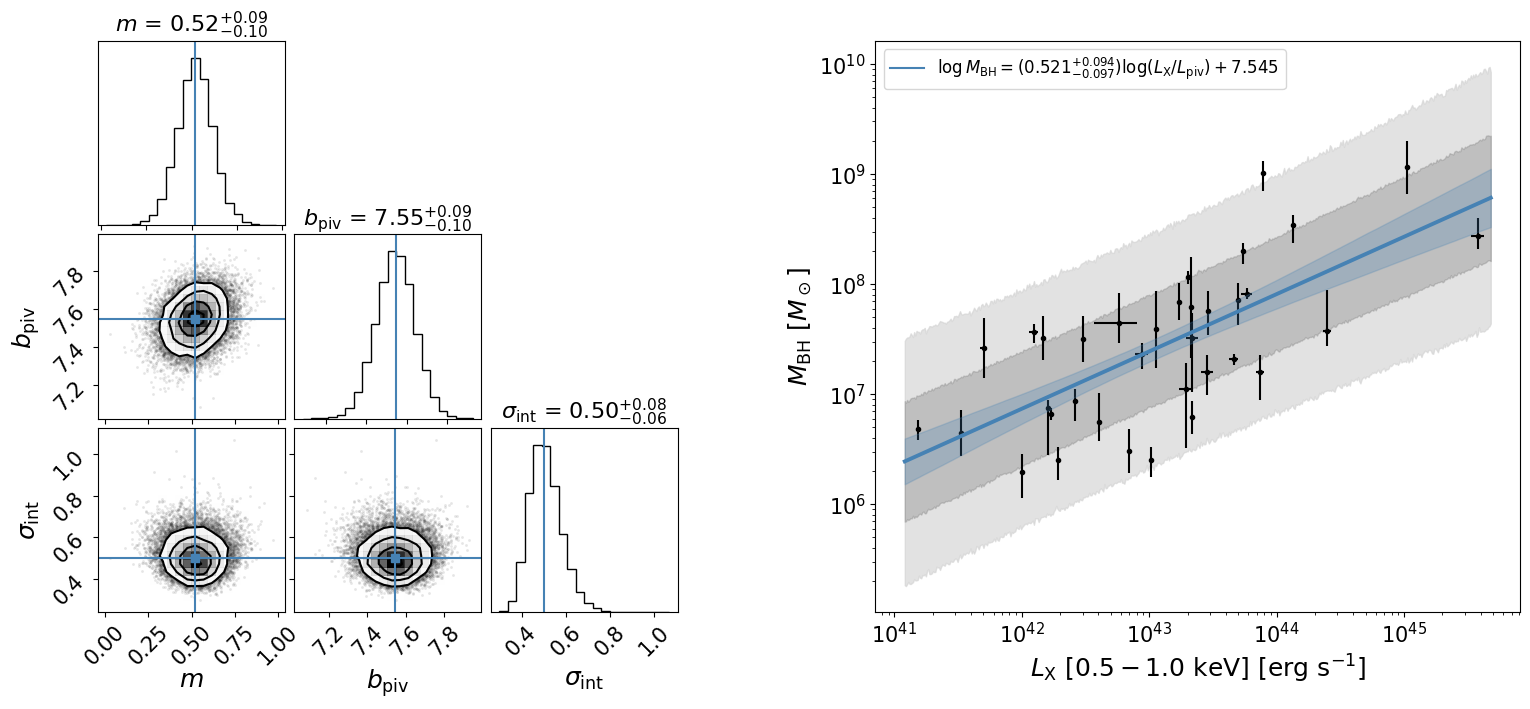}
\caption{$M_\mathrm{BH}$ derived from RM plotted against $L_\mathrm{X}$ measured using eROSITA data in 0.5-1.0 keV energy band. The blue line represents the best-fitting relation with its $1\sigma$ uncertainty, while the grey-shaded regions indicate the 1 and 2$\sigma$ scatter. ($L_\mathrm{piv}=2\times10^{43} \ \mathrm{erg \ s^{-1}}$)}
\label{fig:LX_MBH_eRASS3}
\end{figure}

We find a positive correlation between $L_\mathrm{X}$ and $M_\mathrm{BH}$, with a Pearson correlation coefficient of $r=0.687$ (p-value$=5.270\times10^{-6}$). A MC analysis accounting for measurement uncertainties gave a median correlation coefficient of $r=0.661$, with 68\% and 95\% intervals of $[0.626, 0.695]$ and $[0.589, 0.724]$, respectively. The scatter on the scaling relation is $\sigma_\mathrm{int}=0.50^{+0.08}_{-0.06}$. The best-fitting relation and posterior distributions of the fitted parameters are shown in Figure~\ref{fig:LX_MBH_eRASS3}.\\
 
\section{Discussion and Future Work}

We have confirmed the preliminary finding in \citealt{Mayers2018} that there is a positive correlation between black hole mass and X-ray luminosity. 
Our result is more likely to be representative of the Type-1 AGN  population as a whole than theirs: not only is the sample size larger (35 compared to 30, with an overlap of only 5), we have not filtered for sources with a low or negative $\sigma^2_\mathrm{NXS}$, and the AGN are drawn from a significantly more homogeneous X-ray source superset (derived from eRASS rather than from XMM-Newton pointed observations). Despite the improvements in the sample, the correlation coefficient for our sample is slightly lower than for theirs: 
a median value of $r=0.661$, with a 68\% interval of $[0.626, 0.695]$, compared\footnote{For statistical comparisons with \citealt{Mayers2018}, we have applied the same MCMC analysis method in Section~\ref{sec:method} to their sample.} to $r=0.852$, with a 68\% interval of $[0.834, 0.868]$. Moreover, the scatter on the scaling relation, $\sigma_\mathrm{int}=0.50^{+0.08}_{-0.06}$ is also larger than theirs $\sigma_\mathrm{int}=0.36^{+0.06}_{-0.05}$). It is noteworthy that the measured scatter, of $\simeq 0.5$ dex, in the $M_\mathrm{BH} - L_\mathrm{ X}$ relation is similar to that reported in the well established $M_\mathrm{BH}$ to stellar mass relation ($M_{\rm stellar}$), e.g. \cite{Reines2015}.

The physical mechanisms underlying the measured correlation are unclear. The simplest interpretation is that both the black hole mass and the luminosity are linked to the mass accretion rate. However, this explanation would require the rate at the epoch of observation (i.e. that which drives the luminosity) to be the same as the historically averaged rate (which is responsible for building up the central black hole). This is not the case in practice, as indicated by the fact that the Eddington ratio is known to vary widely across the AGN population, e.g. \cite{Trump2009} measure Eddington ratios of $\lambda_{\rm Edd}=0.1-1$ in a sample of 182 Type-1 AGN (where
$\lambda_\mathrm{Edd}=\frac{L_\mathrm{bol}}{L_\mathrm{Edd}}$, $L_\mathrm{bol}$ is the bolometric luminosity, and $L_\mathrm{Edd}$ is the Eddington luminosity.)
Moreover, the X-ray flux from a given AGN is variable, with times scales ranging from hours through years. This can be caused by a change in the Eddington ratio and/or the line of sight column density. The most extreme examples are the so-called Changing-look AGN Some AGN change their X-ray luminosity output by orders of magnitude over the hours or days, see \cite{Ricci2022} for a review. Hence one would only expect a tight correlation between $M_\mathrm{BH}$ and $L_\mathrm{X}$ if Changing-look AGN were exceptions rather than the norm, and if the distribution of Eddington ratios, despite being wide, was steeply peaked. There is currently no consensus on the latter point: while some studies report a pronounced peak (e.g. \cite{Kollmeier2006}), others find no such feature (e.g. \cite{He2023}, \cite{Lusso2012}, \cite{Woo2002}).

The  current sample size (35) limits the statistical robustness and generalisability of our results. Expanding the sample would allow a more robust assessment of the correlation and its intrinsic scatter. The inclusion of more AGNs at higher redshifts would determine whether the relationship remains applicable across different cosmic epochs. 
%The number of RM $M_\mathrm{BH}$ measurements continues to grow. 
There are two avenues available to expand the sample. First, the number of RM-measured $M_\mathrm{BH}$ estimates is steadily increasing, e.g. the \citealt{Bentz2015} compilation is scheduled for a full update in autumn 2026. Second, eRASS data currently available cover only half of the sky (the half ``owned'' by the German eROSITA Consortium). When data from the other (Russian) half of the sky are made available, the sample size will increase by at least 40\% (because there are 25 AGN in the \citealt{Mayers2018} study that are in that half of the sky). 

The original motivation for this study was to find a low cost (in terms of observing time) method to predict the black hole mass of Type-1 AGN. Unfortunately, the scatter on the scaling relation remains too high ($\sigma_\mathrm{int}=0.50^{+0.08}_{-0.06}$) to make this a practical method to predict $M_{BH}$. That said, our findings can be used to improve mock X-ray skies \citep{comparat2019active} generated from large volume hydrodynamical cosmological simulations. Such mocks are a necessary ingredient for cluster cosmology \citep{Ghirardini2024} and preparation for future missions such as NewAthena \citep{Covas2025}. 

\bibliographystyle{aasjournalv7}
\bibliography{ref}
\end{document}